\documentclass[journal]{IEEEtran}
\usepackage{amsmath,amssymb,amsfonts,graphicx,booktabs,multirow,array,cite,algorithm,algpseudocode,xcolor,tikz,placeins}
\usetikzlibrary{arrows.meta,positioning,calc,fit,shapes.geometric}
\usepackage{dblfloatfix}
\usepackage{cuted}
\usepackage{capt-of}
\usepackage[hidelinks]{hyperref}
\usepackage{float}
\hypersetup{
    colorlinks=false,
    pdfborder={0 0 1},
    citebordercolor={0 1 0},
    linkbordercolor={1 0 0},
    urlbordercolor={1 1 1}
}

\begin{document}
\title{Resource-Conditioned Federated Policy Aggregation for Queue--Energy Coupled Maritime IoT}
\author{Yuli~Liu%
\thanks{Y. Liu is with the Department of Ocean Science and Technology, University of Macau, Taipa, Macau, China (e-mail: mc45259@um.edu.mo).}%
}
\maketitle
\begin{abstract}
Maritime Internet of Things (IoT) sensing couples information delivery with shared uplink capacity, finite batteries, and sensing-induced backlog. These constraints can favor clients with high delivered information even when their model updates yield low joint policy value. We establish a feasible contribution ranking reversal and a shared-bandwidth externality, and develop resource-conditioned fleet aggregation (RCFA). The server evaluates the uniform aggregate and individual proximal policy proposals through paired fleet rollouts stratified by resource regime. Finite-candidate analysis bounds selection uncertainty and characterizes the variance of paired scores. In ten-seed short- and long-budget experiments, RCFA improves normalized discounted return over a longer-rollout FedProx control by 0.03855 and 0.05159, respectively, under identical simulator-slot and model-payload budgets (Holm-adjusted $p=0.0117$). Relative to same-target independent and identically distributed (IID) evaluation, the short-budget gain is 0.00276 (adjusted $p=0.0313$), and the independent long-budget gain is 0.00163 (95\% confidence interval: $[-0.00039,0.00364]$). Both resource regimes are covered in all 520 RCFA rounds. In 50-vessel transfer tests with fixed total bandwidth, short-budget RCFA improves return by 0.07972 over IID (adjusted $p=0.0391$). Weather-conditioned replay using buoy records from the National Oceanic and Atmospheric Administration (NOAA) and matched-budget computational profiling further characterize its performance and cost.
\end{abstract}
\section{Introduction}\label{sec:introduction}
Maritime Internet of Things (IoT) sensing supports offshore and coastal monitoring, where the timely delivery of informative measurements, data backlog, and onboard energy determine sensing mission quality.
Hybrid satellite--aerial--terrestrial networks extend maritime coverage, and unmanned aerial vehicle (UAV)-assisted collection provides access to dispersed marine sensors \cite{r01,r02,fu2024}.
Recent maritime sensing and communication architectures coordinate aerial collection with constrained radio and energy resources \cite{li2025men,li2026monitor}.
In remote waters, fluctuating uplink conditions and limited device energy constrain the service available to newly generated measurements.
Accumulated data can occupy finite buffers, while battery depletion can interrupt information delivery during a mission.
A cooperative UAV--unmanned surface vehicle (USV) architecture places sensing and local control at USVs and coordinates uplink service and information fusion at the UAV.
Shared radio resources and a common fusion objective thereby couple the physical decisions of the entire fleet.

Each USV generates measurements through sensing, compresses data from its existing queue, and uploads the resulting packets through its assigned bandwidth share.
Longer sensing creates additional arrivals and reduces the time available for transmission, while stronger compression increases raw-bit service at the expense of recovered information and processing energy.
Transmit power and bandwidth determine the achievable service; the resulting queue and battery states influence subsequent decisions.
These sequential interactions motivate shared policy learning through proximal policy optimization (PPO) and federated proximal regularization \cite{r17,r15}.
Information-weighted aggregation offers an intuitive way to prioritize local proposals according to the effective information delivered during training.
However, this score reflects both the behavior policy and the client's physical resources: a favorable channel or an ample battery can increase measured contribution independently of the joint value of the updated policy.
A proposal that raises one USV's bandwidth request also changes the service available to its peers, and the nonlinear fusion objective combines their delivered information with diminishing marginal utility.
The central challenge is therefore to identify local policy proposals that improve the fleet objective under coupled queue, energy, and shared-bandwidth dynamics, using a finite server-side evaluation budget.

Existing research addresses this setting through maritime resource control, wireless federated learning (FL), and federated reinforcement learning (FedRL).
Maritime computation and resource-management studies optimize offloading and energy use \cite{qi2025,wang2024marine}, while federated multiagent approaches coordinate trajectories and resource allocation in satellite--UAV marine networks \cite{tesfaw2025}.
Wireless FL jointly designs communication and learning under constrained resources \cite{r04,r05}, and heterogeneous-environment FedRL addresses client-specific dynamics and policy drift \cite{jin2022,wang2024}.
Recent contribution-aware policy aggregation for UAV edge computing prioritizes heterogeneous agents for age-of-information objectives \cite{li2026influence}.
Difficulty- and improvement-aware aggregation (FedDIA) further measures local policy improvement relative to a shared model \cite{feddia2026}.
In the coupled maritime system considered here, a proposal's deployment value depends on the actions of the full fleet, including competition for bandwidth and the evolution of shared utility and future backlogs.
This structure motivates a joint deployment-value criterion, complete-fleet evaluation under a common resource distribution, and an evaluation design that covers each declared resource regime within the available budget.

We propose resource-conditioned fleet aggregation (RCFA) for queue--energy coupled maritime IoT sensing.
USVs generate local proximal PPO proposals using locally retained trajectories, while the UAV coordinates physical actions and a resettable training-time simulator evaluates candidate fleet policies.
The cooperative control objective balances recovered information against energy consumption and backlog, and a battery action shield ensures feasibility within the modeled energy budget.
RCFA constructs a finite candidate set containing the uniform aggregate and interpolations toward individual local proposals.
The server deploys each candidate across the simulated fleet under the same exogenous episode seeds within each prespecified resource regime, and scores its normalized discounted return under the resource mixture.
This paired, stratified design captures cross-client consequences and ensures representation of each resource regime with a positive evaluation allocation.
The selected candidate becomes the next shared model, preserving the federated model-exchange schedule.
Finite-candidate analysis characterizes the selection error and the coverage--variance tradeoff, connecting evaluation allocation to the reliability of candidate comparisons.

The main contributions are threefold:
\begin{itemize}
\item \textbf{Physical coupling and contribution misalignment:} We develop a queue--energy coupled maritime sensing model that jointly represents sensing, compression, shared uplink service, and nonlinear information fusion. Two feasible constructions establish a strict contribution ranking reversal and a shared-bandwidth externality, identifying the physical basis for joint policy evaluation (Secs.~\ref{sec:model}--\ref{sec:misalignment}).
\item \textbf{Resource-conditioned aggregation and analysis:} We formulate finite joint-policy selection over proximal PPO proposals and develop paired, resource-stratified fleet evaluation. We bound finite-candidate selection error and characterize the resource-coverage and paired-score variance properties under the prescribed sampling assumptions (Secs.~\ref{sec:method} and~\ref{sec:theory}).
\item \textbf{Mechanism and system-level validation:} We evaluate frozen local updates, independent ten-seed learning cohorts, interaction- and payload-matched controls, fleet-size transfer, and weather-conditioned replay using National Oceanic and Atmospheric Administration (NOAA) records. RCFA improves normalized discounted return over longer-rollout FedProx by 0.03855 and 0.05159 at matched short and long budgets, respectively, and covers both resource regimes in all 520 RCFA training rounds (Sec.~\ref{sec:setup}).
\end{itemize}
Sections~\ref{sec:related}--\ref{sec:conclusion} present related work, the system and objective, the ranking problem, RCFA, its analysis, evaluation, and conclusions.

\section{Related Work}\label{sec:related}
We review maritime resource control and federated policy learning, corresponding to the physical coupling and aggregation alternatives in our problem.
\subsection{Maritime Sensing and Resource Control}
Maritime UAV collection and hybrid satellite--terrestrial coverage address connectivity for remote devices \cite{r01,fu2024}.
Wind-aware trajectories improve energy efficiency \cite{r02}; cooperative maritime data collection uses relay-based deep reinforcement learning \cite{fu2024}. Maritime integrated sensing and communication and joint sensing, communication, and powering coordinate physical resources \cite{li2025men,li2026monitor}. Cooperative offloading and energy-harvesting mobile edge computing address resource allocation in maritime networks \cite{qi2025,wang2024marine}. Multiagent wireless control and UAV-assisted computing extend sequential decisions to interacting devices \cite{r10,r12,r13}.
These approaches motivate our physical model with fixed vehicle roles and scalar time-correlated channels. Shared uplink allocation and information fusion give each model displacement a fleet-wide effect.
\subsection{Federated Learning and Sequential Control}
Wireless FL jointly manages learning, communication, and data heterogeneity \cite{r04,r05,r07}. Robust learning, partial model aggregation, and adaptive edge FL address unreliable links and resource-constrained participation or computation \cite{r08,chen2024partial,wang2019adaptive}. FedAvg \cite{r16}, FedProx \cite{r15}, and PPO \cite{r17} provide practical building blocks for local policy optimization and shared-model updates. SCAFFOLD corrects client drift through control variates \cite{karimireddy2020}; FedNova normalizes unequal local update lengths \cite{wang2020nova}; FedOpt adapts server-side optimization \cite{reddi2021}.
Heterogeneous-environment FedRL addresses local drift and adaptive aggregation \cite{jin2022,wang2024}. FedSARSA characterizes finite-time on-policy learning under environment heterogeneity \cite{zhang2024sarsa}; FedNPG-ADMM reduces the communication cost of natural policy gradients \cite{lan2023npg}; federated natural policy gradient and actor--critic methods study multi-task policy learning \cite{yang2024nac}. Wireless FedRL supports resource allocation, uplink control, and UAV trajectory fairness \cite{ji2024,zhou2026,tarekegn2025}, as well as UAV-assisted medical IoT, offloading, and freshness-aware collection \cite{seid2023,chen2025racomu,zhang2025aoi}. Tesfaw et al. \cite{tesfaw2025} study trajectory and resource optimization in a satellite--UAV marine network. Li et al. \cite{li2026influence} use contribution-aware policy aggregation for UAV edge computing with an age-of-information objective.
The FedDIA preprint \cite{feddia2026} uses local improvement relative to a common model and difficulty-dependent weighting. FedLAW learns weights on server proxy data \cite{li2023fedlaw}; FedAA controls aggregation weights through validation rewards \cite{he2025fedaa}. Offline FedRL combines estimated returns and policy inconsistency \cite{xu2026offline}.
RCFA targets marginal joint policy value under sensing-induced backlog, battery depletion, shared bandwidth, and nonlinear fusion. The strict reversal in Sec.~\ref{sec:misalignment} identifies a limitation of delivered information as a contribution proxy. A maritime adaptation of FedDIA tests a local-score alternative; same-target independent and identically distributed (IID) joint evaluation isolates the sampling design.
RCFA evaluates candidate policies on complete fleet trajectories under a shared resource mixture. Stratified sampling ensures resource coverage under positive allocations, and paired scenarios support direct candidate comparisons. Optional independent validation follows high-confidence policy improvement principles \cite{thomas2015}.

\section{System Model}\label{sec:model}
We define the fleet, its physical transitions, and its local observations before formulating the shared return.
\subsection{Roles, Timing, and Assumptions}
\begin{figure}[!t]
    \centering
    \includegraphics[width=\columnwidth]{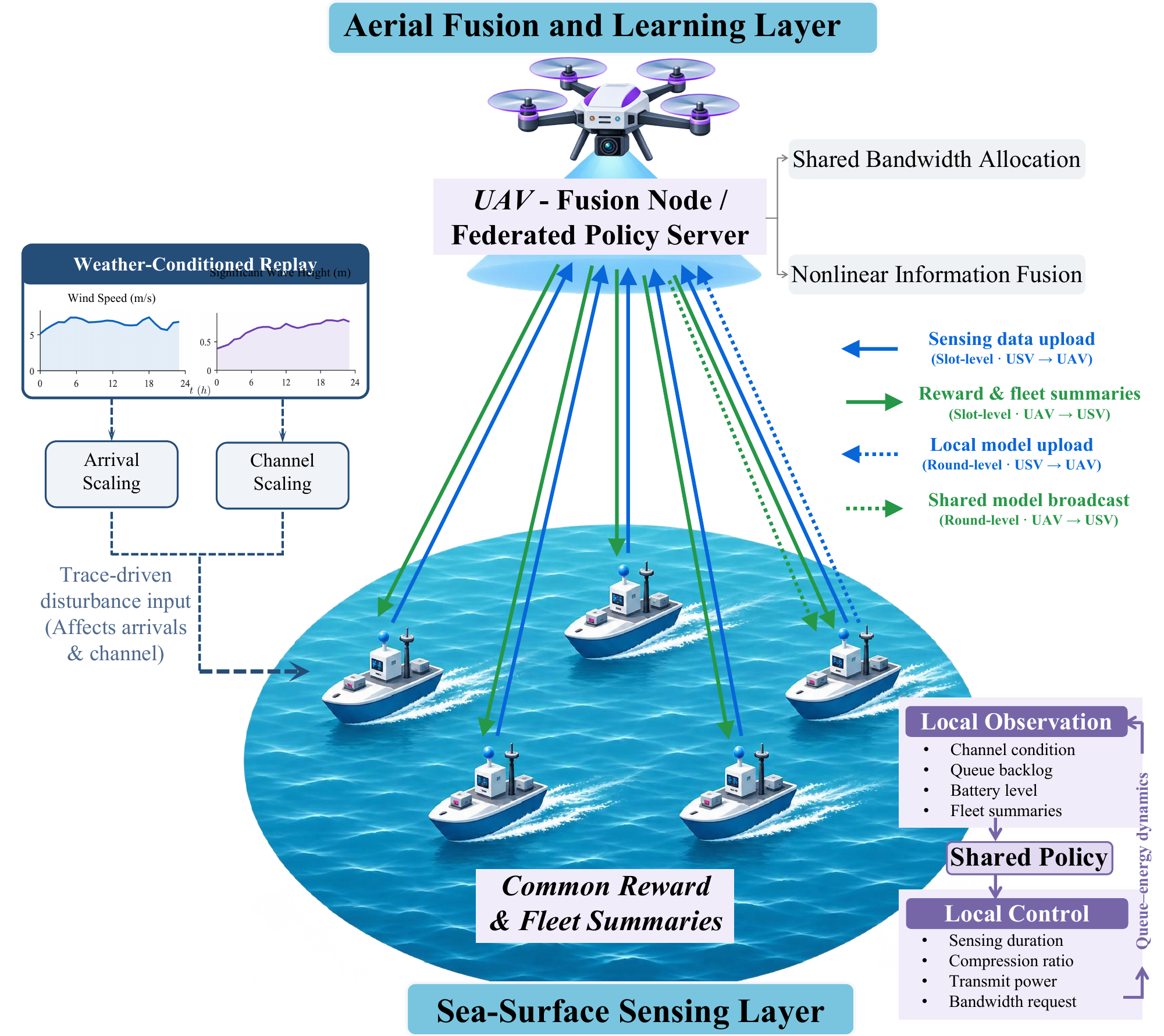}
    \caption{UAV--USV cooperative maritime IoT sensing architecture
    with queue--energy coupled control, shared bandwidth allocation,
    and nonlinear information fusion.
    Weather-conditioned disturbances are used for replay-based evaluation.}
    \label{fig:system}
\end{figure}

Fig.~\ref{fig:system} illustrates the physical loop. We consider one UAV fusion/federated server and $N$ USVs indexed by $\mathcal{N}=\{1,\ldots,N\}$. The time horizon is $t=0,\ldots,T_h-1$, with slot length $\Delta$. USVs observe local channel, backlog and battery; the UAV normalizes bandwidth requests and broadcasts two aggregate statistics. We assume synchronous training, scalar time-correlated gains and fixed vehicle roles. The server has a resettable fleet simulator for training-time evaluation. We use $[x]_a^b=\min\{\max\{x,a\},b\}$ and $[x]_+=\max\{x,0\}$.

\subsection{Action Mapping and Bandwidth Coordination}
A client emits a bounded four-component action
\begin{equation}\label{eq:action}\mathbf a_i(t)=[a_{i,1},a_{i,2},a_{i,3},a_{i,4}]^\top\in[-1,1]^4.
\end{equation}
The corresponding physical decisions are
\begin{subequations}\label{eq:map}\begin{align}
p_i(t)&=\tfrac{a_{i,1}(t)+1}{2}P_{\max},\\
\tau_i(t)&=\tfrac{a_{i,2}(t)+1}{2}s_{\max}\Delta,\\
c_i(t)&=c_{\min}+\tfrac{a_{i,3}(t)+1}{2}(1-c_{\min}),\\
W_i(t)&=\frac{\exp(\eta a_{i,4}(t))}{\sum_{j=1}^N\exp(\eta a_{j,4}(t))}W_{\rm tot}.
\end{align}\end{subequations}
Here $P_{\max}$ is maximum power, $s_{\max}\in(0,1)$ limits sensing duration, $c_{\min}>0$ is the strongest compression level, $W_{\rm tot}$ is shared uplink bandwidth, and $\eta>0$ controls request competition. Consequently,
\begin{equation}\label{eq:feasible}\begin{gathered}
0\le p_i\le P_{\max},\quad 0\le\tau_i\le s_{\max}\Delta,\quad c_{\min}\le c_i\le1,\\
W_i>0,\qquad \sum_{i=1}^N W_i=W_{\rm tot}.
\end{gathered}\end{equation}
Joint softmax normalization is performed at the UAV; raw rollout trajectories remain local. After allocating bandwidth, each USV shields power and sensing time before execution (Appendix~\ref{app:energy_audit}).

\subsection{Channel and Upload Service}
The scalar uplink gain follows a first-order autoregressive (AR(1)) process,
\begin{equation}\label{eq:channel}\begin{split}
g_i(t+1)=\max\{&\mu_i+\rho_g[g_i(t)-\mu_i]\\
&+\sigma_i\sqrt{1-\rho_g^2}\,\epsilon_i(t),g_{\min}\},
\end{split}\end{equation}
where $\epsilon_i(t)\sim\mathcal N(0,1)$, $\mu_i$ and $\sigma_i$ vary by client, $0\le\rho_g<1$, and $g_{\min}>0$. The transmission interval is
\begin{equation}\label{eq:txduration}t_i^{\rm tx}(t)=\max\{\Delta-\tau_i(t),\varepsilon_t\}.
\end{equation}
For noise density $N_0$ and stabilizer $\varepsilon_n$, define
\begin{equation}\label{eq:snr}\Gamma_i(t)=\frac{p_i(t)g_i(t)}{N_0W_i(t)+\varepsilon_n}.
\end{equation}
The volume served from the \emph{pre-existing raw-data queue} is
\begin{equation}\label{eq:service}S_i(t)=\min\left\{Q_i(t),\frac{W_i(t)t_i^{\rm tx}(t)\log_2[1+\Gamma_i(t)]}{c_i(t)}\right\}.
\end{equation}
Smaller $c_i$ serves more raw bits but incurs information loss. Newly arriving bits are available only in later slots.

\subsection{Queue, Information, Battery, and Heterogeneity}
Sensing creates arrivals
\begin{equation}\label{eq:arrivals}\begin{gathered}
A_i(t)=[A_{i,0}+\kappa_i\tau_i(t)+\xi_i(t)]_+,\\
\xi_i(t)\sim\mathcal N(0,\sigma_A^2),
\end{gathered}\end{equation}
where $A_{i,0}$ is a baseline load and $\kappa_i$ converts sensing time to raw bits. The queue transition is
\begin{equation}\label{eq:queue}Q_i(t+1)=[Q_i(t)-S_i(t)+A_i(t)]_0^{Q_{\max}}.
\end{equation}
Compression-dependent recovered information and fused utility are
\begin{align}
I_i(t)&=S_i(t)\exp\!\left[-\beta\left(\frac{1}{c_i(t)}-1\right)\right],\label{eq:information}\\
U(t)&=a_u\log\!\left[1+\vartheta\sum_{i=1}^N I_i(t)\right],\label{eq:utility}
\end{align}
where $\beta\ge0$ controls information loss and $a_u,\vartheta$ define the fused-utility scale. The logarithm models diminishing marginal information utility. Per-slot energy and battery transition are
\begin{align}
E_i(t)&=p_i(t)t_i^{\rm tx}(t)+P_{\rm sense}\tau_i(t)\nonumber\\
&\quad +k_cS_i(t)\left[\frac{1}{c_i(t)}-1\right],\label{eq:energy}\\
B_i(t+1)&=[B_i(t)-E_i(t)]_0^{B_{i,\max}}.\label{eq:battery}
\end{align}
The modeled battery budget covers radio data service, sensing, and compression. Propulsion and model exchange use separate, unmodeled energy budgets. The shield limits expenditure before battery clipping. Bounded factors model non-IID clients:
\begin{equation}\label{eq:hetero}\begin{split}
A_{i,0}&=A_0\zeta_i^A,\quad\kappa_i=\kappa_0\zeta_i^A,\\
\mu_i&=\bar g\zeta_i^G,\quad \sigma_i=\sigma_g\zeta_i^G,\\
B_{i,\max}&=B_{\max}\zeta_i^B.
\end{split}\end{equation}
The load, channel and battery factors are clipped to $[0.2,2.5]$, $[0.2,3]$, and $[0.4,1.6]$. At reset, queue and battery fractions are uniform on $[0.05,0.25]$ and $[0.6,1]$; gain is initialized by the clipped client Gaussian.

\subsection{Local Observations and Model-to-Objective Bridge}
The local policy observes (with negligible $10^{-12}$ denominator stabilizers in the implementation)
\begin{equation}\label{eq:obs}\mathbf o_i(t)=\left[\frac{g_i(t)}{\mu_i},\frac{Q_i(t)}{Q_{\max}},\frac{B_i(t)}{B_{i,\max}},\bar Q(t),\bar g_o(t)\right]^\top,
\end{equation}
where $\bar Q(t)=N^{-1}\sum_j Q_j(t)/Q_{\max}$ and $\bar g_o(t)=N^{-1}\sum_jg_j(t)/\mu_j$ are broadcast summaries. These summaries distinguish network conditions from the local state. Equations~\eqref{eq:map}--\eqref{eq:obs} close the sensing--service--energy feedback loop.

\section{Problem Formulation}\label{sec:problem}
\subsection{Reward and Explicit Feasibility Scope}
We maximize delivered utility while penalizing energy and post-decision backlog. All clients receive the common reward
\begin{equation}\label{eq:reward}
r(t)=U(t)-\lambda_E\frac{\sum_{i=1}^N E_i(t)}{E_{\rm ref}}-\lambda_Q\frac{\sum_{i=1}^N Q_i(t+1)}{Q_{\max}}.
\end{equation}
Here $E_{\rm ref}=1$ J normalizes energy and $\lambda_E,\lambda_Q\ge0$ weight energy and backlog. The queue term represents delay pressure through a soft backlog penalty.

The implemented simulation feasible set is defined over the stacked physical-control vector $\mathbf u(t)=\{p_i(t),\tau_i(t),c_i(t),W_i(t)\}_{i=1}^N$:
\begin{equation}\label{eq:C}\begin{aligned}
\mathcal C(t)=\big\{\mathbf u(t):\; &0\le p_i\le P_{\max},\;0\le\tau_i\le s_{\max}\Delta,\\
&c_{\min}\le c_i\le1,\; W_i>0,\;\sum_jW_j=W_{\rm tot},\\
&0\le Q_i\le Q_{\max},\;0\le B_i\le B_{i,\max}\big\},
\end{aligned}\end{equation}
The action mapping, clipped states and shield enforce these domains and $E_i(t)\le B_i(t)$ up to numerical precision. PPO likelihoods refer to proposed actions, with shielding part of the transition.

\subsection{Cooperative Sequential Control}
Let $\mathcal X$ contain all channel, queue, battery, and heterogeneity states and let $\mathcal O_i$ denote local observation spaces. We specify a decentralized partially observable Markov decision process (Dec-POMDP):
\begin{equation}\label{eq:dec_pomdp}\mathcal M=(\mathcal N,\mathcal X,\{\mathcal A_i\},\mathsf P,\{\mathcal O_i\},r,\gamma),
\end{equation}
with $\mathcal A_i=[-1,1]^4$, transitions $\mathsf P$ induced by~\eqref{eq:channel}--\eqref{eq:battery}, reward~\eqref{eq:reward}, and discount $0<\gamma\le1$. A shared stochastic actor $\pi_\omega$ is executed on distinct local observations. The target is
\begin{equation}\label{eq:objective}\max_{\pi_\omega}J(\pi_\omega)=\mathbb E_{\pi_\omega}\!\left[\sum_{t=0}^{T_h-1}\gamma^t r(t)\right],
\end{equation}
under simulated dynamics, set~\eqref{eq:C}, and the additional modeled energy constraint $E_i(t)\le B_i(t)$ in the shielded environment. The expectation includes disturbances and stochastic actions. Shared bandwidth couples actions across USVs. A slot-wise objective omits the future queue and battery consequences of current actions; stochastic transitions and partial observations motivate sequential cooperative control. The method below combines local stochastic policy optimization with finite deterministic aggregate selection.

\section{When Physical Contribution Misranks Policy Updates}\label{sec:misalignment}
For round $k$, let $m=|\mathcal S_k|$, $\omega_i$ be the local model after the same prescribed PPO update, and $\omega_u=m^{-1}\sum_i\omega_i$ be the uniform reference. The information-based score is
\begin{equation}\label{eq:info_score}
 C_i^{\rm info}=L_i^{-1}\sum_{t=1}^{L_i} I_i(t).
\end{equation}
The score records delivered information under the behavior policy. The information-based weights are $q_i^{\rm info}\propto L_i(\alpha+C_i^{\rm info}/(\bar C+\varepsilon_c))$, where $\bar C=m^{-1}\sum_i C_i^{\rm info}$ and $\alpha>0$.

Let $h\in\{1,\ldots,H\}$ index predeclared fleet resource regimes with nonnegative probabilities $p_h$ summing to one. Each regime specifies a reset distribution and transition law for channels, arrivals, and batteries of the \emph{whole} fleet. For a deterministic deployed policy $\pi_\omega^{\rm d}(\mathbf o)=\tanh\mathbf m_\omega(\mathbf o)$ and the shielded environment, define
\begin{align}\label{eq:target}
 F_h(\omega)&=\mathbb E_h\!\left[\frac{\sum_{t=0}^{T_h-1}\gamma^t r(t)}{\sum_{t=0}^{T_h-1}\gamma^t}\right],\\
 F(\omega)&=\sum_h p_hF_h(\omega).\label{eq:mixturetarget}
\end{align}
Equation~\eqref{eq:mixturetarget} defines the fixed-horizon deterministic deployment counterpart of the stochastic objective~\eqref{eq:objective}. The resource distribution is fixed before scoring; all candidates share the execution convention.

For a preselected interpolation $0<\eta_a\leq1$, define the finite marginal aggregate value
\begin{equation}\label{eq:marginal}
 M_i(\eta_a)=F(\omega_u+\eta_a(\omega_i-\omega_u))-F(\omega_u).
\end{equation}
This finite displacement score measures the joint value of increasing one proposal's share. For differentiable $F$, $M_i(\eta_a)/\eta_a\to\langle\nabla F(\omega_u),\omega_i-\omega_u\rangle$ as $\eta_a\downarrow0$. Finite probes also apply to nondifferentiable shielded transitions. A rank reversal occurs when $(C_i^{\rm info}-C_j^{\rm info})(M_i-M_j)<0$; ties are recorded separately. Because one bandwidth request changes every allocated share, $M_i$ captures cross-client effects beyond local improvement.

\subsection{A Feasible Resource-Induced Reversal}
\textbf{Proposition 1 (existence).} There exist shield-feasible local proposals in the physical model for which the higher-information client has strictly lower marginal aggregate value, and information weighting is worse than uniform weighting.

\emph{Proof:} Consider a one-slot two-USV restriction with $W_i=0.5$ MHz, $\Delta=0.2$ s, $c_i=1$, $\tau_i=0$, and $N_0=10^{-12}$ W/Hz (use the same $10^{-12}$ W denominator stabilizer as the simulator). Let $Q_A=10^5$, $Q_B=10^6$ bits, $g_A=2\times10^{-6}$, $g_B=2\times10^{-7}$, and both batteries equal 1 J. Use the source reward normalization $E_{\rm ref}=1$ J, $a_u=1$, $\vartheta=10^{-6}$, and $\lambda_E=\lambda_Q=0.1$. Fix identical exogenous arrivals for all counterfactuals within the nonclipped region of both queues. A context-conditioned two-parameter policy maps the two clients' distinct observations to their respective powers. The feasible local proposals have power vectors $(0.7,0.3)$ and $(0.3,0.7)$ W. Uniform averaging gives $(0.5,0.5)$ W. Under the common behavior policy with these reference powers, A serves $10^5$ bits and B about $2.63\times10^4$ bits; thus $C_A^{\rm info}>C_B^{\rm info}$. With $\eta_a=1/2$, the two probes give $(0.6,0.4)$ and $(0.4,0.6)$ W. A remains queue-limited for either probe, while B remains link-limited. Total energy is fixed at $0.2$ J. Increasing B's served bits strictly increases reward: in the nonclipped queue region the derivative with respect to its service is
\begin{equation}\label{eq:service_derivative}
 \frac{a_u\vartheta}{1+\vartheta(S_A+S_B)}+\frac{\lambda_Q}{Q_{\max}}>0.
\end{equation}
Consequently $M_A<0<M_B$. For equal rollout lengths and $\alpha>0$, the information-based rule assigns $q_A^{\rm info}>1/2$, shifting power from B to A. Throughout the interval from uniform to that mixture A remains saturated and B remains link-limited; reward strictly falls. The construction concerns feasible proposals of an affine context-conditioned policy; PPO-generated proposals are examined experimentally. \hfill$\square$

\textbf{Proposition 2 (bandwidth externality).} A shield-feasible bandwidth-request increase can improve its issuing client's local service score and reduce joint fleet return. The effect vanishes under fixed bandwidth in the same construction.

\emph{Proof:} Take two USVs with equal powers 0.5 W, queues $10^6$ bits, batteries 1 J, $c_i=1$, $\tau_i=0$, and gains $g_A=2\times10^{-7}<g_B=2\times10^{-6}$. The total bandwidth is 1 MHz. For $z=N_0W+10^{-12}$, $x=Pg$, the service derivative at a queue-unlimited link is $\Delta[\log(1+x/z)-WN_0x/(z(z+x))]/\log 2$. It is positive; its derivative with respect to $x$ is $\Delta(x+10^{-12})/[(z+x)^2\log 2]>0$. At equal bandwidth, increasing A's request therefore increases its own service while decreasing total service. Energy remains fixed; the joint utility and queue penalty are strictly monotone in total service in the nonclipped region. For an A request change from 0 to 0.2 (B fixed at 0), the source simulator gives service increments of $+379$ and $-13{,}704$ bits, local A reward increment $0.000374$, and joint reward increment $-0.011444$. Fixed equal bandwidth yields identical scores. This action-level construction holds for feasible context-conditioned policies. \hfill$\square$

The constructions motivate the frozen-update diagnostics in Sec.~\ref{sec:budget}.

\section{Resource-Conditioned Fleet Aggregation}\label{sec:method}
\begin{figure}[!t]
    \centering
    \includegraphics[width=\columnwidth]{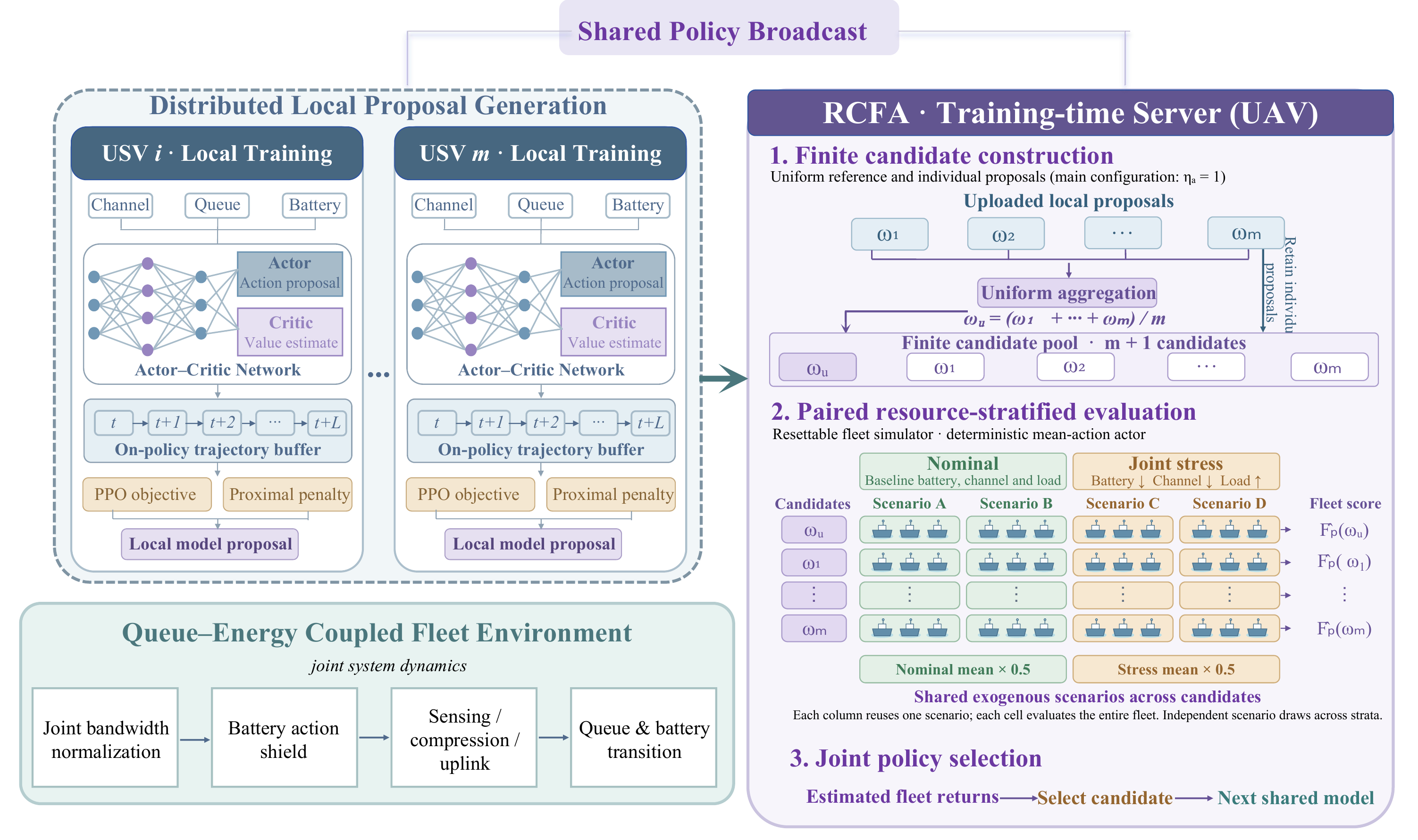}
    \caption{Architecture of resource-conditioned fleet aggregation
    (RCFA), comprising distributed local policy optimization,
    finite candidate construction, paired resource-stratified
    fleet evaluation, and joint policy selection.
    The resource mixture is fixed before selection,
    and raw trajectories remain local.
    The candidate pool illustrates the setting $\eta_a=1$.}
    \label{fig:rcfa}
\end{figure}

Fig.~\ref{fig:rcfa} illustrates the two time scales of RCFA: USVs interact with the physical environment during local rollouts, and the server evaluates and selects candidate shared models at each federated round. Algorithm~\ref{alg:rcfa} summarizes the resulting update procedure.
\subsection{Shared Controller and Local Proposal Generator}
The shared actor uses local observations~\eqref{eq:obs} and a Gaussian latent action with tanh squashing:
\begin{equation}\label{eq:policy}
 \mathbf x_i\sim\mathcal N(\mathbf m_\omega(\mathbf o_i),\operatorname{diag}(\boldsymbol\sigma_\omega^2)),\qquad \mathbf a_i=\tanh\mathbf x_i.
\end{equation}
The tanh network uses a linear actor mean and a critic predicting common reward-to-go. Local trajectories stay at the USV. Advantages are computed using generalized advantage estimation (GAE). The proximal PPO loss is
\begin{align}
 \rho_t(\omega)&=\pi_\omega(\mathbf a_t\mid\mathbf o_t)/\pi_{\omega_{\rm old}}(\mathbf a_t\mid\mathbf o_t),\\
 \mathcal L_i(\omega)&=-\mathbb E_t\min\{\rho_t\hat A_t,\operatorname{clip}(\rho_t,1-\epsilon,1+\epsilon)\hat A_t\}\nonumber\\
 &\quad+c_v\mathbb E_t(V_\omega-\hat G_t)^2-c_e\mathbb E_t\mathcal H(\pi_\omega)\nonumber\\
 &\quad+\tfrac{\lambda_{\rm prox}}2\|\omega-\omega^{(k)}\|_2^2.\label{eq:prox}
\end{align}
Aggregation covers the actor--critic parameters, and deterministic validation evaluates the shared actor across all USVs.

\subsection{One Optimization Target for Aggregation}
The server receives $m$ model displacements and forms the uniform model $\omega_u$. For equal rollout lengths, let $u_i=1/m$ and let $e_i$ be the $i$th simplex vertex. The predeclared candidate set is
\begin{align}
 \mathcal Q_{\eta_a}&=\{u\}\cup\{(1-\eta_a)u+\eta_a e_i:i\in\mathcal S_k\},\label{eq:candidates}\\
 \omega(q)&=\sum_{i\in\mathcal S_k}q_i\omega_i.\label{eq:aggregate}
\end{align}
Selection uses this finite subset of the simplex and requires forward evaluations. The displacement from $\omega_u$ is bounded by $\eta_a\max_i\|\omega_i-\omega_u\|_2$. The final configuration sets $\eta_a=1$, giving the uniform model and $m$ individual local proposals. A single selected model is broadcast to the fleet at the next round.

For each $h$, draw $n_h$ independent complete fleet episodes and reuse their exogenous random seeds for all candidates. Denote the normalized discounted return of candidate $q$ by $Y_{h,j}(q)$. The primary selection rule is
\begin{align}
 \widehat F_P(q)&=\sum_h\frac{p_h}{n_h}\sum_{j=1}^{n_h}Y_{h,j}(q),\\
 \widehat q&=\arg\max_{q\in\mathcal Q_{\eta_a}}\widehat F_P(q).\label{eq:select}
\end{align}
A fixed candidate order resolves ties, starting with the uniform model. Appendix~\ref{app:gate} describes optional independent validation of a selected candidate. Paired scores measure the entire candidate under common constraints, including bandwidth externalities and future queues and batteries.

\begin{algorithm}[!t]
\caption{Resource-conditioned fleet aggregation}\label{alg:rcfa}
\begin{algorithmic}[1]
\Require Shared model, resource mixture $(p_h)$, $\eta_a$, evaluation budgets $n_h$.
\For{federated round $k=1,\ldots,K$}
 \State Broadcast current model; collect shielded local trajectories.
 \State Obtain participating proposals by~\eqref{eq:prox}.
 \State Form $\mathcal Q_{\eta_a}$ and candidates by~\eqref{eq:candidates}--\eqref{eq:aggregate}.
 \State Evaluate candidates on paired joint pilot episodes.
 \State Select $\widehat q$ by~\eqref{eq:select}.
 \State $\omega^{(k+1)}\gets\omega(\widehat q)$.
\EndFor
\State \Return shared actor.
\end{algorithmic}
\end{algorithm}

\subsection{Information Access and Complexity}
For $P$ parameters, $m$ participants and $b_p$ bits/parameter, model exchange costs approximately $2mPb_p$ bits per round with participant-only downloads. Server-side simulation retains the model exchange schedule and adds $(m+1)\sum_h n_h T_h$ joint simulated slots. Candidate construction admits $O(mP)$ arithmetic by reusing the uniform model; the released weighted-sum implementation constructs each of the $m+1$ mixtures separately in $O(m^2P)$ arithmetic. Candidate storage is $O(mP)$. Each slot evaluates all $N$ actors and shared dynamics. Online inference uses the shared actor; field evaluation additionally requires radio service, resets and energy accounting.

\section{Finite-Candidate and Resource-Coverage Analysis}\label{sec:theory}
Assume that, conditional on the training history and fixed candidates, the episodes within each resource stratum are independent across seed draws and strata. Common random numbers within a paired comparison are allowed. Assume $Y_{h,j}(q)\in[a_h,b_h]$ for all candidates, and set $R_h=b_h-a_h$. For the physical model, with nonnegative utility parameters and $\beta\geq0$, a valid conservative interval is
\begin{align}
 b_h&=a_u\log(1+\vartheta NQ_{\max}),\\
 a_h&=-\lambda_E N E_{\max}/E_{\rm ref}-\lambda_Q N,\\
 E_{\max}&=P_{\max}\Delta+P_{\rm sense}s_{\max}\Delta\\
 &\quad+k_cQ_{\max}(c_{\min}^{-1}-1).\nonumber
\end{align}
Since $0\leq I_i\leq S_i\leq Q_{\max}$, the normalized discounted return lies in this conservative interval even with unbounded disturbances.

\textbf{Theorem 1 (finite candidate selection).} Let $M=|\mathcal Q_{\eta_a}|\leq m+1$, $V_P=\sum_h p_h^2 R_h^2/n_h$, and $\epsilon_P=\sqrt{V_P\log(2M/\delta_P)/2}$. With probability at least $1-\delta_P$,
\begin{equation}\label{eq:regret}
 F(\omega(\widehat q))\geq\max_{q\in\mathcal Q_{\eta_a}}F(\omega(q))-2\epsilon_P.
\end{equation}
Uniform is in the candidate set, giving the relative lower bound $F(\omega_u)-2\epsilon_P$.

\emph{Proof:} For fixed $q$, each weighted observation has range length $p_h R_h/n_h$. Hoeffding's inequality gives $\Pr(|\widehat F_P(q)-F(\omega(q))|>\epsilon)\leq2\exp(-2\epsilon^2/V_P)$. A union bound over the fixed set yields simultaneous errors at most $\epsilon_P$. On this event, comparing the empirical maximizer to the true finite-set maximizer incurs at most two such errors. \hfill$\square$

\subsection{Coverage and the Allocation Tradeoff}
For two regimes, a stress probability $p$ and $B$ IID draws, the probability of no stress observation is $(1-p)^B$; the probability of omitting either regime is $(1-p)^B+p^B$. A fixed stratified design with at least one draw in each regime makes this probability zero. This coverage property is deterministic conditional on the designated regime labels. For a fixed candidate with regime means $\mu_h$ and variances $\sigma_h^2$, independent sampling gives
\begin{align}
\operatorname{Var}(\widehat F_{\rm strat})&=\sum_h p_h^2\sigma_h^2/n_h,\\
\operatorname{Var}(\widehat F_{\rm IID})&=\frac{\sum_h p_h\sigma_h^2+\sum_h p_h(\mu_h-\mu)^2}{B},
\end{align}
where $\mu=\sum_h p_h\mu_h$ and $B=\sum_h n_h$. Under exact proportional allocation $n_h=Bp_h$, stratification removes the between-regime variance term for that candidate. Equal allocation matches the balanced two-regime target and overrepresents a rare stress regime. Overallocating draws to a rare regime can increase within-regime estimation variance relative to proportional allocation. Candidate selection depends on score differences. For candidates $q,q'$, let $Z_h=Y_h(q)-Y_h(q')$ under a shared episode seed, with mean $d_h$ and variance $s_h^2$. For independent draws across regimes, the variance gap between IID and fixed stratification is
\begin{equation}\label{eq:pairvariance}
\begin{aligned}
&\operatorname{Var}(\widehat d_{\rm IID})-\operatorname{Var}(\widehat d_{\rm strat})\\
&=\frac{\sum_h p_h(d_h-\bar d)^2}{B}+\sum_h\left(\frac{p_h}{B}-\frac{p_h^2}{n_h}\right)s_h^2.
\end{aligned}
\end{equation}
where $\bar d=\sum_h p_hd_h$. Common random numbers (CRN) give $s_h^2=\operatorname{Var}(Y_h(q))+\operatorname{Var}(Y_h(q'))-2\operatorname{Cov}(Y_h(q),Y_h(q'))$. Proportional allocation removes between-regime variation in paired differences. Equal allocation under unequal probabilities can increase paired variance if the allocation penalty exceeds the between-regime term. For the balanced target and $B=4$, RCFA sets $n_0=n_1=2$; IID has an omission probability of $1/8$ per round. Section~\ref{sec:budget} examines ranking accuracy and coverage.

\section{Simulation and Evaluation}\label{sec:setup}
We evaluate whether joint fleet selection improves the shared controller at matched communication and simulation budgets, and whether resource stratification improves selection over IID evaluation.
The experiments cover learning performance, budget allocation, frozen-update mechanisms, resource perturbations, fleet-size transfer, and weather-conditioned replay.
\subsection{Simulation Setup and Benchmarks}
The five-USV, one-UAV simulator follows Table~\ref{tab:params}; each policy operates on local observations and the same battery shield.
The target distribution assigns equal probability to nominal and joint-stress environments. Joint stress halves battery capacity and channel mean and standard deviation, and multiplies arrival baseline, slope and noise standard deviation by 1.5. Local PPO collects nominal trajectories; candidate scoring and final tests use deterministic mean actions. Each candidate episode spans 200 slots.
\begin{figure}[!t]
    \centering
    \includegraphics[width=0.49\columnwidth]{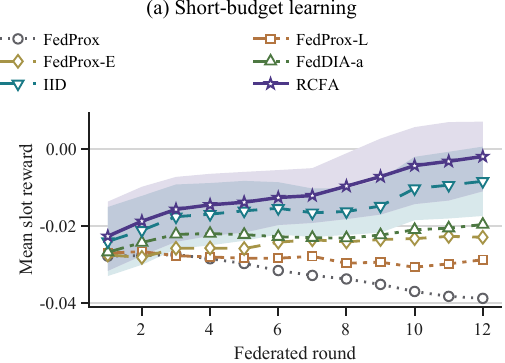}
    \hfill
    \includegraphics[width=0.49\columnwidth]{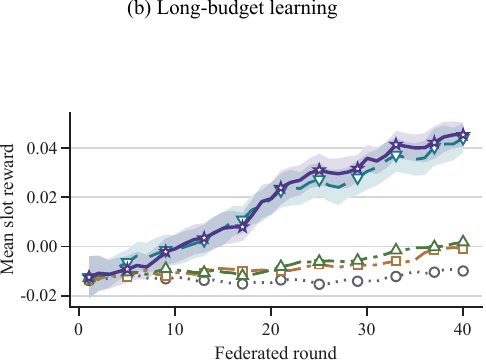}
    \caption{Learning performance under (a) 12 rounds with
    256 local slots and (b) 40 rounds with 1024 local slots.
    The nominal monitoring metric is mean slot reward;
    the curves average ten seeds. Shaded bands show seed
    standard errors for RCFA and IID.}
    \label{fig:learning}
\end{figure}
\begin{table}[!t]\centering\caption{Principal Simulation and Learning Parameters}\label{tab:params}\footnotesize
\resizebox{\columnwidth}{!}{\begin{tabular}{@{}ll@{}}\toprule Parameter & Value\\\midrule
$N,\Delta,T_h$ & $5,0.2$ s, $200$ slots\\
$W_{\rm tot},N_0,P_{\max}$ & 1 MHz, $10^{-12}$ W/Hz, 1 W\\
$B_{\max},Q_{\max}$ & 10 J, $10^7$ bits\\
$A_0,\kappa_0,\sigma_A$ & $2\times10^4,6\times10^5,2\times10^4$ (bits, bits/s, bits)\\
$\bar g,\sigma_g,\rho_g$ & $10^{-6},6\times10^{-7},0.9$\\
$\operatorname{sd}(\zeta^A,\zeta^G,\zeta^B)$ & $(0.35,0.40,0.15)$ before clipping\\
$s_{\max},c_{\min},\beta,\eta$ & $0.6,0.15,1,2$\\
$P_{\rm sense},k_c$ & 0.3 W, $10^{-9}$ J/bit\\
$a_u,\vartheta,\lambda_E,\lambda_Q$ & $1,10^{-6},0.1,0.1$\\
$\gamma$, GAE, PPO clip & $0.99,0.95,0.2$\\
Hidden layers; Adam learning rate & $128,128$; $3\times10^{-4}$\\
PPO epochs; minibatch & 4; 256\\
$\lambda_{\rm prox},\eta_a$ & $0.005,1$\\
Participation; strata & Full; nominal/stress $p_h=0.5$\\
Short budget & 12 rounds $\times$ 256 slots; 10 seeds\\
Long budget & 40 rounds $\times$ 1024 slots; 10 seeds\\
Pilot episodes/candidate & 4; $n_0=n_1=2$\\\bottomrule
\end{tabular}}
\end{table}

\begin{figure*}[!t]
    \centering
    \includegraphics[width=0.325\textwidth]{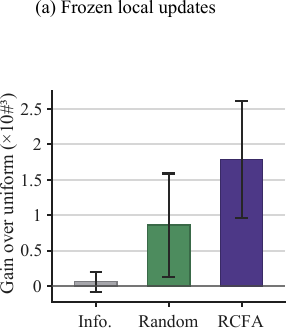}
    \hfill
    \includegraphics[width=0.325\textwidth]{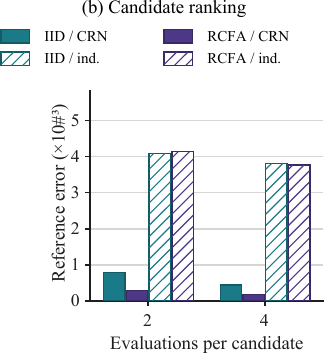}
    \hfill
    \includegraphics[width=0.325\textwidth]{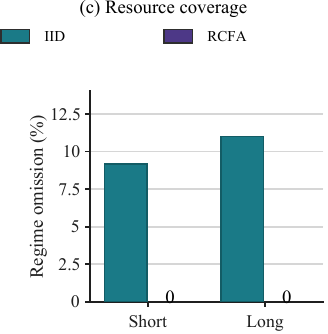}
    \caption{Aggregation diagnostics.
    (a) Historical frozen-update gains over uniform
    ($\eta_a=0.5$, six checkpoints; unadjusted 95\% intervals).
    (b) Reference selection error from 20 frozen
    expanded-candidate pools ($\eta_a=1$, $B=2,4$),
    with common random numbers (CRN) and independent seeds.
    (c) Regime omission in final training
    ($\eta_a=1$, $B=4$; 120 and 400 rounds).}
    \label{fig:mechanism}
\end{figure*}
FedProx uses uniform aggregation with proximal updates. FedProx-L converts the 4800 joint evaluation slots per round into a longer continuous local rollout, matching the total simulation and model exchange budgets of RCFA. IID uses the same six candidates, resource target, episode count and shared random-number convention, with IID regime labels. FedDIA-a adapts the local improvement and difficulty weights in \cite{feddia2026}: positive improvement and inverse pre-update return are min--max normalized; the product $\widetilde\Delta_i(1+\widetilde d_i)$ enters a unit-temperature softmax, with uniform fallback for nonpositive improvements. A common pre-model and one-client replacements yield 24 local evaluation episodes per round. The adaptation uses maritime PPO and the same physical model. An additional short-budget control, FedProx-E, retains the original rollout and uses 24 independent 200-slot training episodes split across the two resource regimes, with bootstrap at truncated boundaries. This sensitivity control tests a different allocation of the same training interaction budget.

The primary metric is normalized discounted return~\eqref{eq:mixturetarget}; fleet backlog, fusion utility, energy and overflow characterize the physical tradeoff. The short and long cohorts use ten independent seeds each (202611400--409 and 202611410--419). Python, NumPy and PyTorch generators are seeded; PyTorch uses deterministic algorithms. Within each RCFA/IID cohort, initial candidate tensors coincide exactly. The updated comparator runs share seed numbers and test seeds; cross-archive first-round differences reach $3.27\times10^{-6}$, and 24 same-host repeated candidate hashes coincide. Final tests use eight episodes per regime for short training and sixteen for long training. Statistics use seed means, unadjusted paired Student-$t$ confidence intervals (CIs) and two-sided exact sign-flip tests under null sign symmetry. Holm correction is applied separately to family A (four short-budget contrasts: RCFA versus IID at $B=2,4$, lagged allocation versus IID and RCFA at $B=4$), family B (RCFA versus FedProx, FedProx-L and FedDIA-a at both budgets; six contrasts), and family C (RCFA versus IID across six resource settings, three fleet sizes and NOAA, at both budgets; twenty contrasts). The long-budget IID test and exploratory FedProx-E test are reported separately. The archive lists all 32 tests; all completed seeds enter the analyses.

\subsection{Learning and Overall Performance}
Fig.~\ref{fig:learning} plots the nominal monitoring reward versus federated rounds, showing finite-budget learning trajectories. Independent final evaluation in Table~\ref{tab:main} measures performance under the balanced resource target.
RCFA attains returns of $-0.03324$ and $0.00290$ after short and long training, compared with $-0.03600$ and $0.00128$ for IID. Relative to IID, the paired short-budget improvement is 0.00276 (9/10 seeds; Holm-adjusted $p=0.03125$), and the long-budget estimate is 0.00163 (8/10; $p=0.08008$). The latter interval includes zero. RCFA reduces mean fleet backlog by 4.3\% and 1.9\%, respectively, compared with IID. Compared with FedProx and FedDIA-a, its return gains are 0.04992 and 0.02662 in short training and 0.07723 and 0.03942 in long training. All six updated baseline comparisons, including FedProx-L, yield ten positive paired differences and Holm-adjusted $p=0.01172$.
\begin{table*}[!t]\centering\caption{Independent final performance and paired return effects (ten seeds per method and budget).}\label{tab:main}\footnotesize
\begin{tabular}{@{}lrrrrrrr@{}}\toprule
Method & Return $\uparrow$ & Queue (Mbit) $\downarrow$ & Utility $\uparrow$ & Overflow (Mbit) $\downarrow$ & RCFA gain & 95\% paired CI & $p$\\
\midrule\multicolumn{8}{@{}l}{\emph{Short budget}}\\
RCFA & \textbf{-0.03324} & \textbf{11.784} & \textbf{0.04797} & \textbf{0.209} & -- & -- & --\\
IID & -0.03600 & 12.314 & 0.04757 & 0.235 & 0.00276 & [0.00039, 0.00513] & 0.03125\\
FedDIA-a & -0.05986 & 13.830 & 0.04252 & 0.585 & 0.02662 & [0.01983, 0.03341] & 0.01172\\
FedProx & -0.08316 & 16.377 & 0.04260 & 1.153 & 0.04992 & [0.03466, 0.06518] & 0.01172\\
FedProx-L & -0.07179 & 15.055 & 0.04231 & 0.813 & 0.03855 & [0.02632, 0.05078] & 0.01172\\
FedProx-E & -0.05607 & 12.523 & 0.03746 & 0.402 & 0.02283 & [0.01525, 0.03041] & 0.00195$^\dagger$\\
\midrule\multicolumn{8}{@{}l}{\emph{Long budget}}\\
RCFA & \textbf{0.00290} & \textbf{9.122} & \textbf{0.07030} & \textbf{0.094} & -- & -- & --\\
IID & 0.00128 & 9.302 & 0.06949 & 0.107 & 0.00163 & [-0.00039, 0.00364] & 0.08008\\
FedDIA-a & -0.03652 & 9.966 & 0.04147 & 0.242 & 0.03942 & [0.03151, 0.04734] & 0.01172\\
FedProx & -0.07433 & 14.727 & 0.04298 & 1.155 & 0.07723 & [0.06802, 0.08645] & 0.01172\\
FedProx-L & -0.04869 & 10.511 & 0.03697 & 0.426 & 0.05159 & [0.04075, 0.06243] & 0.01172\\
\bottomrule
\end{tabular}\par\vspace{2pt}
\begin{minipage}{.97\textwidth}\footnotesize
Bold marks RCFA. CI denotes confidence interval; intervals are unadjusted. Short-budget IID $p$ uses Holm family A (four tests), and FedProx, FedProx-L and FedDIA-a use family B (six tests). Long-budget IID $p$ is raw. $^\dagger$Exploratory raw $p$ for FedProx-E.
\end{minipage}\end{table*}

Uniform FedProx uses the local proposals equally; FedDIA-a ranks local pre/post improvements. Joint candidate scores account for the common queue and fusion objective across the fleet. The controlled diagnostics below examine this relationship at fixed local updates.
\subsection{Aggregation Mechanisms and Evaluation Allocation}\label{sec:budget}
The two feasible examples in Sec.~\ref{sec:misalignment} motivate a frozen-update comparison. In a historical six-checkpoint study with five resource settings and 150 local proposals, information and marginal-value rankings disagree in 46.7\% of untied pairs (checkpoint-level 95\% CI: 34.2--59.1\%). Joint evaluation gains 0.001785 over uniform, compared with 0.000061 for information weighting and 0.000861 for random directions (Fig.~\ref{fig:mechanism}(a)). These historical diagnostics use the contracted set ($\eta_a=0.5$) to examine contribution--value misalignment. Both contrasts with joint selection have Holm-adjusted sign-flip $p=0.0625$; reference scores are finite-pool estimates.

A second historical six-checkpoint intervention ($\eta_a=0.5$) holds proposals and episode seeds fixed while changing physical couplings. Joint versus local-proxy gains are 0.000527 under full coupling, 0.000683 with fixed bandwidth, 0.000561 with separable fusion, 0.000707 with both changes, and 0.000020 with reference-state replay. These contrasts suggest temporal state coupling as a useful diagnostic; the intervention contrasts have Holm-adjusted $p>0.05$. The bandwidth construction in Proposition 2 identifies an explicit action-level externality independently of the average intervention estimates.

For the final candidate set ($\eta_a=1$: uniform and individual proposals), Fig.~\ref{fig:mechanism}(b) compares selection error on 20 frozen expanded-candidate pools with separate selection and reference samples. At four evaluations per candidate, the mean reference error falls from 0.000433 with IID to 0.000166 with proportional stratification under CRN. Independent candidate seeds yield errors near 0.0038 in both cases. The resampled estimates describe conditional performance on these pools. In the ten-seed short-budget cohort, at each of $B=2$ and $B=4$, RCFA selects the best independent-reference candidate in all ten initial rounds, compared with nine for IID. At $B=2$, deterministic replay of each seed's first selection divergence gives a mean independent-reference gain of 0.001444 for the RCFA selection (8/10 positive; 95\% CI: $[-0.000105,0.002992]$). These results connect paired evaluation to candidate ordering.

At the final four-evaluation budget, IID omits a regime in 11/120 short and 44/400 long rounds, whereas RCFA covers both in every round (Fig.~\ref{fig:mechanism}(c)). At two evaluations per candidate the paired short-budget gain is 0.003308 (adjusted $p=0.0625$). A lagged variance allocation that chooses 1:3, 2:2 or 3:1 using the previous balanced round's paired-score variances improves the four-evaluation fixed design by 0.000156 ($p=0.125$). Fixed 2:2 allocation supplies coverage with a simple schedule, consistent with~\eqref{eq:pairvariance} for the balanced target. Rare-stress targets may favor proportional or variance-sensitive allocations; the variance identity characterizes this tradeoff.

\begin{figure*}[!t]
\centering
    \includegraphics[width=0.325\textwidth]{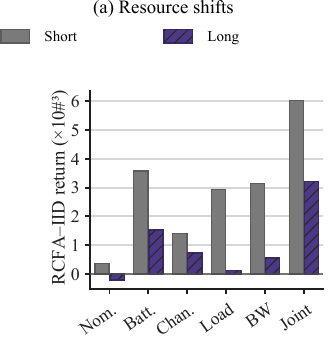}
    \hfill
    \includegraphics[width=0.325\textwidth]{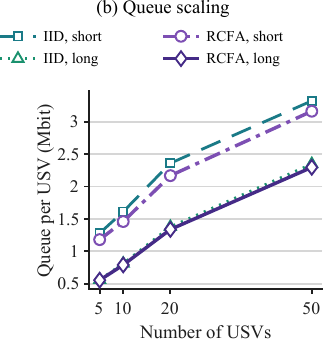}
    \hfill
    \includegraphics[width=0.325\textwidth]{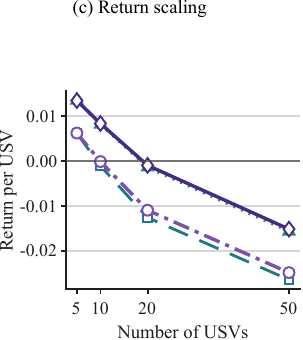}
    \caption{Resource and scale transfer.
    (a) RCFA--IID return differences for discrete resource
    interventions, in units of $10^{-3}$.
    (b),(c) Average per-USV queue and return at fixed total
    bandwidth, with separate short- and long-budget models.
    All policies were trained with five USVs. Nom., Batt., Chan., and BW denote nominal, battery, channel, and bandwidth settings.}
    \label{fig:resource}
\end{figure*}
\subsection{Simulation Budget and Computational Cost}
\begin{figure}[!t]
    \centering
    \includegraphics[width=0.49\columnwidth]{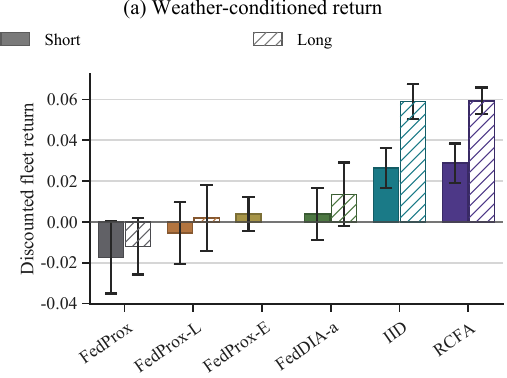}
    \hfill
    \includegraphics[width=0.49\columnwidth]{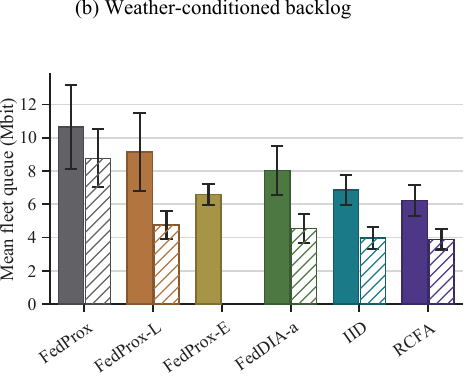}
    \caption{NOAA 2025 replay across nine held-out windows
    and ten seeds per method. (a) Discounted return,
    (b) mean fleet queue. Vertical bars show unadjusted
    95\% intervals of seed-level means; FedProx-E was
    evaluated in short training.}
    \label{fig:noaa}
\end{figure}
Table~\ref{tab:cost} separates local training, candidate evaluation, and monitoring interactions. RCFA uses the same model payload as IID, FedDIA-a, and FedProx-L. Redirecting server evaluation slots to continuous local PPO trajectories yields lower return for FedProx-L by 0.03855 (short) and 0.05159 (long). Longer continuous trajectories can accumulate battery-depleted states. FedProx-E improves the short-budget uniform control by distributing the extra slots across fresh resource-conditioned episodes; RCFA retains a gain of 0.02283 (95\% CI: $[0.01525,0.03041]$, exploratory raw $p=0.00195$). These controls establish the benefit of the tested allocation of a fixed simulator budget to joint policy evaluation.
\begin{table}[!t]\centering\caption{Per-model simulator and payload budgets}\label{tab:cost}\footnotesize
\begin{tabular}{@{}lrrrr@{}}\toprule
Method & Local & Search & Total & Mbit\\\midrule
\multicolumn{5}{@{}l}{\emph{Short budget}}\\
RCFA/IID/FedDIA-a & 3072 & 57600 & 63072 & 135.20\\
FedProx-L/E & 60672 & 0 & 63072 & 135.20\\
FedProx & 3072 & 0 & 5472 & 135.20\\\midrule
\multicolumn{5}{@{}l}{\emph{Long budget}}\\
RCFA/IID/FedDIA-a & 40960 & 192000 & 240960 & 450.68\\
FedProx-L & 232960 & 0 & 240960 & 450.68\\
FedProx & 40960 & 0 & 48960 & 450.68\\\bottomrule
\end{tabular}\par\vspace{2pt}\footnotesize Total includes 2400/8000 monitoring slots; final tests are separate.
\end{table}

Serial profiling on an AMD EPYC 9V74 central processing unit (CPU) uses one core, one computational thread, three repeated three-round trials per method, and 15,768 simulator slots per trial. RCFA takes 3.654 s versus 14.178 s for FedProx-L; candidate evaluation accounts for 2.465 s of the RCFA time. PPO updates account for 0.202 and 3.741 s, respectively. Five-USV batch inference has mean median latency 10.552 and 10.582 $\mu$s; peak memory is 367.17 and 368.87 MiB. This simulator-level profile measures the cost of the two budget allocations on the specified hardware. Radio transmission and simulator calibration have separate deployment costs.

\subsection{Resource Perturbations and Fleet-Size Transfer}
Fig.~\ref{fig:resource}(a) shows the final models under six discrete resource settings. Lower channel gain and shared bandwidth restrict~\eqref{eq:service}; higher load increases~\eqref{eq:arrivals}, and reduced batteries shorten active service. For fixed total bandwidth, increasing the number of USVs reduces the capacity available per vehicle.
Each of the 110 final models is evaluated twice per resource setting and twice at each of 10, 20 and 50 USVs (1980 episodes), using matched exogenous seeds and the same five-USV-trained actors. Short-budget RCFA has positive mean return differences from IID across all six resource settings; long-budget RCFA has positive differences in five. Joint-stress differences are 0.00603 and 0.00321, respectively. At 50 USVs, short-budget RCFA improves return from $-1.32473$ to $-1.24501$, lowers mean per-USV backlog from 3.322 to 3.162 Mbit, and reduces mean overflow from 0.4000 to 0.2159 Mbit. The return difference of 0.07972 has a 95\% interval of $[0.04638,0.11307]$ and Holm-adjusted $p=0.03906$ in family C. The remaining scenario comparisons have adjusted $p>0.05$. Fig.~\ref{fig:resource}(b),(c) uses per-USV metrics to describe scale transfer; raw fleet return differences compare policies within a fixed fleet size.

\subsection{Weather-Conditioned Replay and Energy Feasibility}
Weather-conditioned replay examines performance under time-varying disturbances outside the synthetic training traces.
We process 695,453 raw meteorological records from five NOAA National Data Buoy Center (NDBC) stations (42002, 42012, 42019, 42020, 42085), 2023--2025 \cite{noaa}, remove missing markers, aggregate hourly, and fill isolated one-hour gaps bracketed by valid records. Nonoverlapping 200-hour windows number 13, 9 and 9 in the respective years.
Per-station 2023 wind and wave-height 10th/90th percentiles define clipped normalized variables $w,v$. The stress index $s=0.45w+0.55v$ scales channel gain and arrivals by $1.15-0.45s$ and $0.75+0.8s$, respectively. The underlying AR(1) gain evolves separately from weather scaling.
The 110 final models each replay nine held-out 2025 windows, yielding 990 episodes. Each hourly observation indexes one 0.2-s simulated slot; the temporal compression preserves the order of the weather observations.
Fig.~\ref{fig:noaa} compares return and backlog across the final methods. RCFA attains mean returns of 0.02880 and 0.05934 in short and long training, respectively, compared with 0.02642 and 0.05888 for IID. The corresponding paired differences are 0.00239 and 0.00045, with both adjusted $p=1.0$ in family C. The replay results show positive mean differences under the specified weather mapping.
The mapping supplies a reproducible parametric disturbance model; quantitative field calibration requires concurrent sea-state, radio-link and traffic observations.
The physical model covers sensing, data upload, compression and battery depletion over simulated missions. Vessel propulsion, mobility-dependent channel state information and field radio latency require separate measurements.

Across the final 110 models, 2560 independent final-test, 1980 resource and fleet-size transfer, and 990 weather-conditioned replay episodes have zero substantive unfunded-energy violations at the $10^{-7}$ J tolerance; the largest unfunded energy is $3.49\times10^{-9}$ J. The shield proof in Appendix~\ref{app:energy_audit} establishes energy feasibility in exact arithmetic.
Together, the experiments support joint evaluation at matched interaction and payload budgets, deterministic resource coverage, and improved mean candidate selection under balanced resource conditions. The significant short-budget IID gain and positive independent long-budget estimate characterize the effect of stratification; field calibration and evolving resource distributions remain deployment considerations.

\FloatBarrier
\section{Conclusion}\label{sec:conclusion}
This study investigated federated policy aggregation for maritime IoT sensing under shared bandwidth and finite energy.
The queue--energy model establishes contribution ranking reversal and a shared-bandwidth externality.
The shared objective balances delivered information, energy, and backlog under shielded physical actions.
RCFA selects a shared policy through paired, resource-stratified fleet evaluation, with finite-candidate and variance analyses describing its statistical properties.
Independent short- and long-budget studies yield return gains of 0.03855 and 0.05159 over longer-rollout FedProx at matched simulator and payload budgets. Relative to IID, RCFA achieves a significant short-budget gain, a positive long-budget estimate, and a significant short-budget gain in 50-vessel transfer.
Field calibration of the weather-to-link mapping, propulsion-aware energy accounting, and evaluation under evolving resource distributions are natural extensions.

\appendices
\section{Energy-Feasible Execution}\label{app:energy_audit}
With $d_i(t)=[E_i(t)-B_i(t)]_+$, summation of~\eqref{eq:battery} gives $\sum_{t,i}d_i(t)=\sum_{t,i}E_i(t)-\sum_iB_i(0)+\sum_iB_i(T_h)$. To ensure energy feasibility, upper-bound the proposed action cost by
\begin{equation}\label{eq:safecoeff}\begin{split}
C_i={}&p_i\Delta+P_{\rm sense}\tau_i\\
&+k_c\frac{W_i\Delta}{c_i}\frac{\Gamma_i}{\ln2}(c_i^{-1}-1).
\end{split}\end{equation}
If the action would exceed $B_i$, execute $p_i'=s_i p_i$, $\tau_i'=s_i\tau_i$ with $s_i=\min\{1,B_i/C_i\}$ for $C_i>0$ (a zero-cost action requires no scaling), retaining $c_i,W_i$. Since $t_i^{\rm tx}\leq\Delta$ and $\log_2(1+s_i\Gamma_i)\leq s_i\Gamma_i/\ln2$, each term of~\eqref{eq:energy} is bounded by its counterpart in $s_i C_i$. Thus $E_i'\leq B_i$ in exact arithmetic. Scaling preserves the action domains. The numerical audit uses an unfunded-energy tolerance of $10^{-7}$ J.

\section{Optional Independent Validation}\label{app:gate}
Freeze the pilot choice $\widehat q$ before drawing independent paired validation returns $D_{h,j}=Y_{h,j}(\widehat q)-Y_{h,j}(u)$. With $v_h$ pairs and return ranges $R_h$, set
\begin{equation}\label{eq:gate}
 \widehat d=\sum_h p_h\bar D_h,\quad b=\sqrt{2\log(1/\delta)\sum_h p_h^2R_h^2/v_h}.
\end{equation}
Accept $\widehat q$ if $\widehat d-b\geq-\epsilon_{\rm tol}$; otherwise return $u$.

\textbf{Theorem 2.} Conditional on fixed proposals and the pilot, under Sec.~\ref{sec:theory}'s independence and boundedness assumptions,
\begin{equation}\label{eq:cert}
 \Pr\{F(\omega_{\rm gated})<F(\omega_u)-\epsilon_{\rm tol}\}\leq\delta.
\end{equation}
\emph{Proof:} Each difference lies in $[-R_h,R_h]$. One-sided Hoeffding gives $d\geq\widehat d-b$ with probability at least $1-\delta$; acceptance then meets the threshold and rejection returns the reference. Conditioning and averaging establish the bound. \hfill$\square$

Across decisions, error budgets sum by a union bound. The auxiliary empirical Bernstein radius uses stratum terms $\sqrt{2s_h^2\ell/v_h}+14R_h\ell/[3(v_h-1)]$, with $\ell=\log(2H/\delta)$ \cite{maurer2009}.

\end{document}